\documentclass{aastex631}
\usepackage{subfigure}
\usepackage{threeparttable}
\usepackage{newtxtext,newtxmath}
\usepackage{booktabs}
\usepackage{amsmath}
\usepackage{graphicx}
\usepackage{textcomp}
\usepackage[figuresright]{rotating}
\shorttitle{Investigation of superhumps in SU UMa-type dwarf novae}
\shortauthors{Liu et al.}
\graphicspath{{./}{figures/}}

\begin{document}

\title{Investigation of superhumps in SU UMa-type dwarf novae based on the
	observations of TESS}

\correspondingauthor{Shengbang Qian}
\email{qsb@ynao.ac.cn}

\author[0000-0002-0690-3273]{Wei Liu}
\affiliation{Yunnan Observatories, Chinese Academy of Sciences, PO Box 110, Kunming 650216, China}
\affiliation{Department of Astronomy, Key Laboratory of Astroparticle Physics of Yunnan Province, Yunnan University, Kunming 650091, China}
\affiliation{University of Chinese Academy of Sciences, No.1 Yanqihu East Rd, Huairou District, Beijing, 101408, China}
\affiliation{Key Laboratory of the Structure and Evolution of Celestial Objects, Chinese Academy of Sciences, PO Box 110, Kunming 650216, China}

\author{Shengbang Qian}
\affiliation{Yunnan Observatories, Chinese Academy of Sciences, PO Box 110, Kunming 650216, China}
\affiliation{Department of Astronomy, Key Laboratory of Astroparticle Physics of Yunnan Province,  Yunnan University, Kunming 650091, China}
\affiliation{University of Chinese Academy of Sciences, No.1 Yanqihu East Rd, Huairou District, Beijing, 101408, China}
\affiliation{Key Laboratory of the Structure and Evolution of Celestial Objects, Chinese Academy of Sciences, PO Box 110, Kunming 650216, China}



\begin{abstract}
We report the superhumps analysis of seven SU UMa-type dwarf novae based on the observations of Transiting Exoplanet Survey Satellite (TESS). Superhumps are seen during superoutbursts of SU UMa-type dwarf novae. The month-long data sets of TESS are well suited for studying the variation of superhumps. We selected seven non-eclipsing SU UMa-type dwarf novae with superhumps in which TESS light curves are available and have not yet been studied. The stages A, B, and C of superhumps in superoutbursts were determined by O-C method. The results indicate that not all complete superoutbursts show obvious three stages, such as DT Oct and the second superoutburst of J1730+6247. We calculated the superhump periods for each stage and the mean periods for whole superoutbursts.  Taking the stage A superhump method, the mass ratios (M2/M1) were estimated.  According to the results,  the seven stars are pre-bounce systems with mass ratios ranging from 0.1 to 0.2. By combining the orbital periods and the mean superhump periods, the precession periods were calculated. The results show that the precession periods of the seven SU UMa stars are about 2 days.

\end{abstract}



\section{Introduction}

Cataclysmic variable stars (CVs) are a kind of interacting binary stars which contain a white dwarf and a red dwarf \citep{2003cvs..book.....W}.   The white dwarf accretes mass from the red dwarf through the inner Lagrange point and an accretion disk will form around the white dwarf in non-magnetic CVs.  Dwarf novae are thought of non-magnetic CVs, the main feature of which is unpredictable repeated outbursts with amplitude ranging from 2-5 magnitude.  

The periods of CVs typically range from an hour to a dozen hours.
According to the CVs evolution model, the orbital period gradually decreases from long periods to short periods, and increases as they reach the minimum period ($\sim 75$ min) \citep{2001cvs..book.....H,2003cvs..book.....W}.
Furthermore, the orbital period distribution of CVs shows a gap between 2 to 3 hours \citep{2003A&A...404..301R}.   SU UMa-type dwarf novae (SU UMa stars) with orbital period generally below the period gap are a subclass of dwarf novae. Unlike other dwarf novae, SU UMa stars exhibit frequent outbursts, including normal outbursts and superoutbursts. Superoutbursts are about one magnitude brighter than normal outbursts with longer duration and less frequency. A superoutburst usually triggers by a normal outburst which is also called a precursor outburst.
The  mechanism of superoutbursts is considered of the combination of disk instability and tidal instability, which is different from normal  outbursts \citep{1989PASJ...41.1005O,1974PASJ...26..429O,2001NewAR..45..449L}. The tidal instability of accretion disks is due to a resonance between the orbiting secondary star and outer disk particle orbits with a 3:1 period ratio \citep{1990PASJ...42..135H}. 

Superhumps usually occur during the plateau of superoutbursts and disappear with the end of the superoutburst and will not appear in the precursor outburst. They are periodic brightness variations with a period a few percent longer than the orbital period because of the apsidal precession of the elliptic disk. Therefore, superhump period is related to precession period and orbital period. The relation among precession period($P_{prec}$), superhump period($P_{sh}$) and orbital period($P_{orb}$) is $1/P_{prec} = 1/P_{orb} - 1/P_{sh}$. Superhumps were first discovered by \citet{1974A&A....36..369V} in December 1972, during a superoutburst of the SU UMa star VW Hyi. Since then, they have been confirmed to be present in every SU UMa star undergoing a superoutburst, and have become the defining characteristic of SU UMa stars \citep{2000NewAR..44...45O}.   The period and amplitude of superhumps are not constant but constantly change with the superoutburst. According to the evolution of superhump period, \citet{2009PASJ...61S.395K} divided superhumps into three stages: a longer superhump period in early envolutionary stage (stage A), varying period in the middle stage (stage B), a shorter superhump period in the final stage (stage C). 
Superhumps also had been used to determining binary parameters, the mass ratio can be estimated by using stage A superhump period \citep{2013PASJ...65..115K}.     \citet{2020PASJ...72...14K} made a series of studies on the period variation of superhumps in SU UMa-type stars. But most of their research is based on ground-based telescopes, which are difficult to obtain a complete superoutburst light curve.

The intervals of superoutbursts (supercycle) are tens of days to hundreds of days and superoutbursts are unpredictable, therefore their light curves are not easy to obtain.  The month-long data sets of TESS are well suited for studying the variation of superhumps. With the continuous light curves observed by the Transiting Exoplanet Survey Satellite \citep[TESS;][]{ricker2015transiting}, the opportunity arises for systematic analysis of the evolution of superhumps. 

In this paper, we present a superhumps study of seven SU UMa stars based on TESS data. Section 2 introduces the observations and data reduction of these SU UMa stars. Section 3 shows the analysis of superhumps' evolution during superoutbursts. Based on this, we constrain mass ratios and precession periods.  Section 4 contains a conclusion.

\section{Data and methods}

The main mission of TESS is to search for exoplanets via transit detection, and many superoutbursts of SU UMa stars were observed by the way. Continuous monitoring allows us to study the properties of superhumps better than the ground-based telescope. The light curves were downloaded from the Mikulski Archive for Space Telescopes  (MAST) data archive
\footnote{https://mast.stsci.edu/portal/Mashup/Clients/Mast/Portal.html}. The TESS data used in this work are available at MAST:\dataset[10.17909/1pd2-wk57]{\doi{10.17909/1pd2-wk57}}. TESS divides the sky into 26 sectors,  observing the southern ecliptic hemisphere in the first year of mission operation and the northern ecliptic hemisphere in the second year. 
It observes a sector of sky for about 27 days, corresponding to two spacecraft orbits.  Every sector has a one-day gap for transferring data back.
The detectors attached to TESS are sensitive to the wavelengths ranging from 600nm to 1000nm.
The light curves provided as Simple Aperture Photometry (SAP) with a cadence of 120 seconds were used to analyse the superhumps. The observation of two adjacent sectors is continuous with only a short gap in between, and a light curve may consist of multiple consecutive sector observations.

The Ritter-Kolb Catalog \citep[RKcat;][]{2003A&A...404..301R}\footnote{https://wwwmpa.mpa-garching.mpg.de/RKcat/} contains coordinates, apparent magnitudes, orbital parameters, stellar parameters of the components, and other characteristic properties of CVs with known or suspected orbital periods, together with a comprehensive selection of the relevant literature. Version 7.24 of RKcat was used in this work,  which is the final edition published in 2016 with 1429 CVs.

Using the RKcat cross match TESS observational data, some SU UMa stars with superoutbursts and superhumps were found. We selected seven non-eclipsing stars  whose data had not been studied or published. Without the influence of eclipse, it is more conducive to determining the accurate light maximum times. The details of observations of individual objects are listed in table \ref{tab:log}.  Some of these stars have light curves of a complete superoutburst, and some have more than one.

 \begin{table}
	\centering
	\caption{Journal of observations}
	\label{tab:log}
	\begin{tabular}{lccc} 
		\hline
		Targets & LC number & sector & BJD 2457000+\\
		
		\hline

		V1504 Cyg & 1 & 15 & 1711.4 - 1737.4\\
		& 2 & 40, 41 & 2390.7 - 2446.6\\
		& 3 & 54, 55 & 2769.9 - 2824.3\\
		
		V503 Cyg & 1 & 14, 15 & 1683.4 - 1737.4\\
		& 2 & 41 & 2420.0 - 2446.6\\
		& 3 & 55, 56 & 2797.1 - 2853.1\\
		
		TY Psc & 1 & 17 & 1764.7 - 1789.7\\
		& 2 & 57 & 2853.4 - 2882.1\\
		
		SS UMi & 1 & 14 - 17 & 1683.4 - 1789.7\\
		& 2 & 19 - 23 & 1816.1 - 1954.9\\
		& 3 & 25, 26 & 1983.6 - 2035.1\\
		& 4 & 40, 41 & 2390.7 - 2446.6\\
		& 5 & 47 & 2579.8 - 2606.9\\
		& 6 & 49, 50 & 2637.5 - 2691.5\\
		& 7 & 52-54 & 2718.6 - 2796.1\\
		& 8 & 56, 57 & 2825.3 - 2882.1\\
		& 9 & 60 & 2936.9 - 2962.6\\
		IX Dra & 1 & 14 - 16 & 1683.4 - 1763.3\\
		& 2 & 18, 19 & 1790.7 - 1841.1\\
		& 3 & 21, 22 & 1870.4 - 1926.5\\
		& 4 & 24 - 26 & 1955.8 - 2035.1\\
		& 5 & 40, 41 & 2390.7 - 2446.6\\
		& 6 & 48, 49 & 2607.9 - 2664.3\\
		& 7 & 51 - 53 & 2692.9 - 2769.0\\
		& 8 & 55 - 57 & 2797.1 - 2882.1\\
		& 9 & 60 & 2936.9 - 2962.6\\
	
		DT Oct & 1 & 27 & 2036.3 - 2060.6\\
		& 2 & 39 & 2361.8 - 2389.7\\
		J1730+6247 & 1 & 14 - 20 & 1683.4 - 1868.8\\
		& 2 & 22 - 26 & 1899.3 - 2035.1\\
		& 3 & 40, 41 & 2390.7 - 2446.6\\
		& 4 & 47 - 50 & 2579.8 - 2691.5\\
		& 5 & 52 - 60 & 2718.6 - 2962.6\\
		
		\hline
	\end{tabular}
\end{table}

O-C method is a typical method for studying the periodic variation of signals, where O is the observed special moment, and C represents the calculated moment according to the ephemeris. When the initial epoch is selected, C can be calculated based on the mean superhump period. Using quadratic polynomial fitting method, the light maximum times of each superhump are obtained. The result data of light maximum times and O-C values are available from GitHub\footnote{https://github.com/Liu-Wei-astro/7-non-eclipsing-SU-UMa-stars} and China-VO at doi:\dataset[10.12149/101275]{\doi{10.12149/101275}}. 
Through the O-C analysis, the different stages of superhumps can be determined.

The superoutbursts trend of the light curves are removed by the Locally Weighted Scatterplot Smoothing (LOWESS) method.  By removing the trend, the changes in amplitude of superhumps can be obtained. To search for the periods of each stage in the detrended light curves, the generalized Lomb-Scargle method  \citep{1976Ap&SS..39..447L,1982ApJ...263..835S,1989ApJ...338..277P}
was applied. The frequency errors are calculated by the method proposed by \citet{1999A&A...349..225B}.  

Stage A superhumps were identified to reflect the dynamical precession rate of the disk at the radius of the 3:1 resonance. The stage A superhump method is a dynamical method to determine the mass ratio in that it relies only on celestial mechanics, which makes it superior to older, empirical approaches\citep{2022arXiv220102945K}. 
Using orbital period and stage A superhump period, the fractional superhump excess in frequency is
\begin{equation}
	\epsilon^{\ast} = 1 - \frac{P_{orb}}{P_{shA}}.
\end{equation}
Where $P_{orb}$ is the orbital period, $P_{shA}$ is the period of stage A superhumps.
According to the relationship between mass ratio $q$ and fractional superhump excess $ \epsilon^{\ast}$, 
\begin{equation}
	q = -0.0016+2.60\epsilon^\ast +3.33 (\epsilon^\ast)^2+79.0 (\epsilon^\ast)^3,
	\label{Eq:q}
\end{equation}
given by \citet{2013PASJ...65..115K}, the mass ratios (q = M2/M1) can be derived. Equation \ref{Eq:q} is a polynomial approximation, which has a small  maxmimum error of 0.0004 in q, in the range $0.025 \leq q \leq 0.394 $. The exact equations for computing q from the fractional excess of stage A superhumps can be found in \citet{2013PASJ...65..115K}, subject to the correction in \citet{2016PASJ...68..107K}.

The mean precession period ($P_{prec}$) during a superoutburst can be estimated by the equation 
\begin{equation}
	\frac{1}{P_{prec}} = \frac{1}{P_{orb}} - \frac{1}{P_{sh}},
	\label{Eq:prec}
\end{equation}
where $P_{sh}$ is the mean period of superhumps.

\section{Results}

\subsection{V1504 Cyg}
The supercycle of V1504 Cyg is about 137 days with a superhump period of 0.0722 days \citep{2005ASPC..330..379A}. The V magnitude range ascribed in RKcat is from 18.5 to 13.4.  Its orbital period was determined to be 1.668 hours [0.06951(5) days]  based on the radial velocities analysis\citep{1997PASP..109.1359T}. \citet{2012arXiv1206.6762C} discovered superhumps and negative superhumps in its Kepler light curves and estimated the parameters of this system. Using the same data, \citet{2013PASJ...65...50O} first found that the superhumps appear near the maximum of the precursor outburst and grow smoothly from the precursor to the main superoutburst. The authors also indicated that the superoutburst was initiated by tidal instability (as evidenced by the growing superhump).  Its optical spectra show double-peaked Balmer emission lines, arguing for a moderate to relatively high orbital inclination (perhaps $40^{\circ}$ – $60^{\circ}$) \citep{2013AJ....145..109H}. 

	\begin{table}[h]
		\centering
		\caption{Superhump period in V1504 Cyg}
		\label{tab:V1504cyg}
		\begin{tabular}{lccccc} 
			\hline
			sector & stage & E & BJD 2457000+  & period(d) &Error\\
			\hline
			15 &  A & 0 - 10 & 1730.01 - 1730.75& 0.073074 & 0.000088\\
			&  B & 11 - 69 & 1730.75 - 1735.01& 0.072209 &0.000011\\
			
			41 &  A & 0 - 8 & 2422.72 - 2423.31& 0.073529 &0.000120\\
			&  B & 9 - 66 & 2423.31 - 2427.50 & 0.072240&0.000010\\
			&  C & 67 - 100 & 2427.50 - 2432.46& 0.071912&0.000009\\
			
			\hline
		\end{tabular}
	\end{table}

\begin{figure}[h]
	
	\subfigure[$T_{max} = BJD 2458730.017335+ 0.072214E$]{
		\includegraphics[width=0.5\columnwidth]{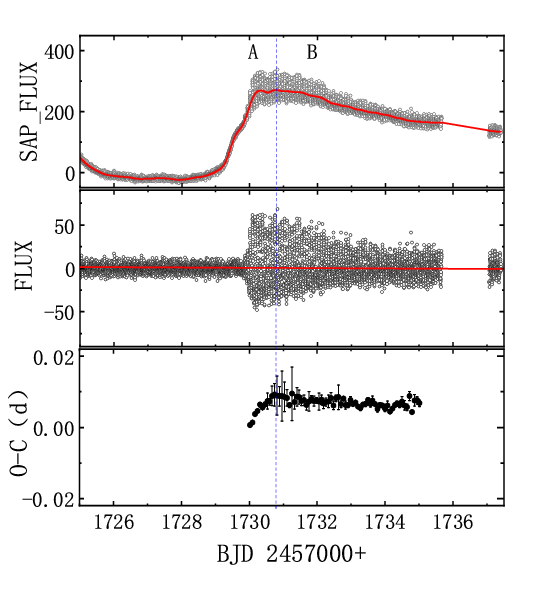} 
	}
	\subfigure[$T_{max} = BJD 2459422.719834+ 0.072214E$]{
		\includegraphics[width=0.5\columnwidth]{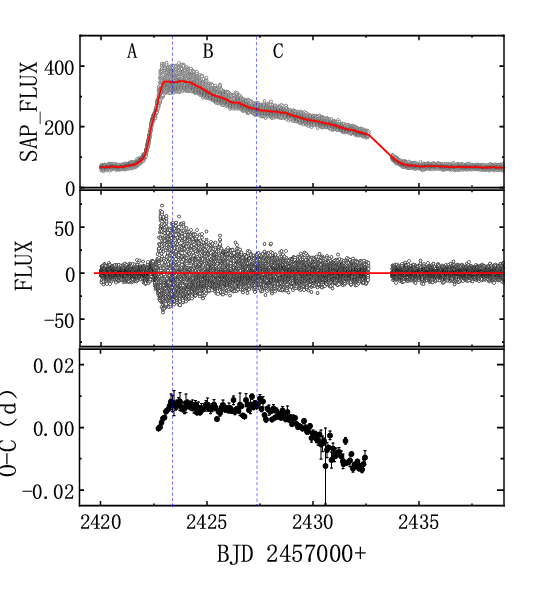} 
	}

	\caption{The two figures correspond to two superoutbursts of V1504 Cyg. Top panels in each figure are light curves of the superoutbursts, the red lines show the trend of the superoutbursts. Middle: the detrended light curves. Bottom: O-C curves of superhumps. Ephemerides of  superhumps for each superoutburst are marked under each figure. The blue dashed lines represent the temporal dividing line among stage A, B and C. stage C of figure (a) superoutburst was not well observed by TESS }\label{fig:V1504cyg}
\end{figure}

Five sector data, including two superoutbursts of this target, were obtained by TESS (see figure \ref{fig:V1504cyg}). The stage C superhumps were not well observed in superoutburst (a). The superoutburst (b) lasts 12.4 days and there is no obvious precursor outburst in the superoutbursts. The O-C
analysis of superhumps at each superoutburst uses different initial
epoch, so the cycle number (E) always starts with epoch 0. The superhump ephemeris are marked under the subfigure of figure \ref{fig:V1504cyg}. The superhumps of different stage are analyzed and the results are summerized in table \ref{tab:V1504cyg}. Taking the average value [0.0733(1) days] of the two stage A superhump periods and the mass ratio of q = 0.153(7) is derived. The mean superhumps period is 0.072214(8) days, and the precession cycle is 1.86(3) days.

\subsection{V503 Cyg}
V503 Cyg shows a short supercycle length (89 days) with a V magnitude range from 18.1 to 13 \citep{2003A&A...404..301R}. It shows frequent normal outbursts with typical recurrence times of 7–9 days, but sometimes the normal outbursts are very infrequent\citep{2002PASJ...54.1029K}. \citet{1995PASP..107..551H} reported a mean superhump period of 0.08101(4) days and detected negative superhumps in quiescence. The orbital period of 0.077760(3) days was discovered by \citet{2012Ap.....55..494P}, and the authors did not find negative superhumps in quiescence and normal outbursts. The superhump period was derived as 0.081084 days according to the 2011 July and October superoutbursts \citep{2013PASJ...65...23K}.  In latter observations, the negative superhump was never seen again.
\begin{table}[h]
    \centering
    \caption{Superhump period in V503 Cyg}
    \label{tab:V503Cyg}
    \begin{tabular}{lccccc} 
        \hline
        sector & stage & E & BJD 2457000+  & period(d) &Error\\
        \hline
        
        15 &  A & 0 - 16  & 1727.00 - 1728.32  & 0.082781&0.000044 \\
        &  B & 17 - 48 & 1728.32 - 1730.84 & 0.081431&0.000025 \\
        &  C & 49 - 83 & 1730.84 - 1733.76 & 0.081175&0.000030\\

        \hline
    \end{tabular}
\end{table}

\begin{figure}[h]
	\centering
	\includegraphics[width=0.6\columnwidth]{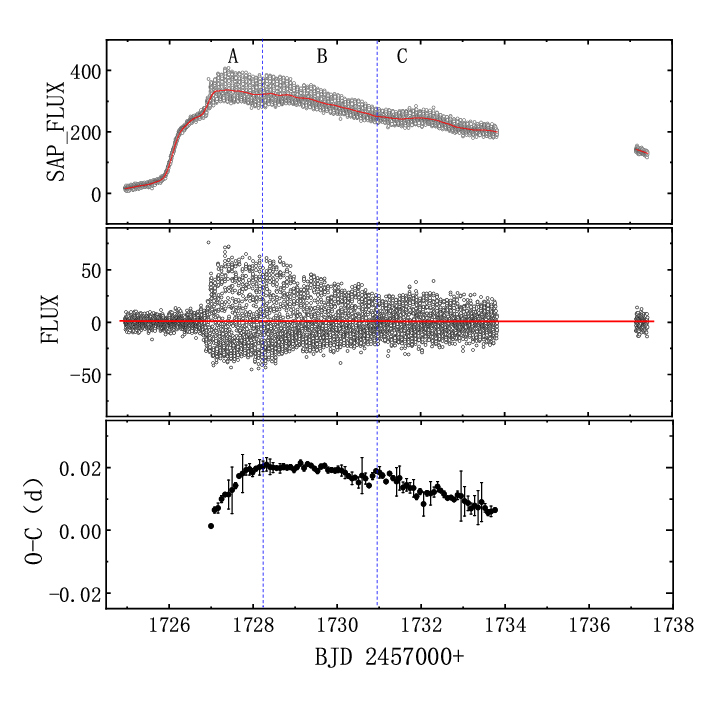}
	\caption{Top: light curves of superoutburst V503 Cyg, the red line shows the trend of the superoutburst. Middle: the detrended light curves. Bottom: O-C diagram of V503 Cyg during BJD 2459069 - 2459086 superoutburst. The ephemeris of $T_{max} = BJD 2458726.9966571+0.081477E$ was used to draw this O-C diagram. The two blue dashed lines represent the temporal dividing line among stage A, B and C.}
	\label{fig:V503Cyg}
\end{figure}

Five sectors data of this target were acquired by TESS, but only one superoutburst in 2019 September is in the data(see figure \ref{fig:V503Cyg}).  A total of 83 cycles of superhumps in all three stages exist in the TESS light curve of this superoutburst.  The superhump ephemeris is marked under figure \ref{fig:V503Cyg}. The O-C diagram shows three stages superhumps in the superoutburst. The information of each stage superhumps was summarized in table \ref{tab:V503Cyg}. Taking the value [0.082781(44) days] of the stage A superhump period and the orbital period, the q = 0.186(2) is derived. Using the mean superhumps period of 0.081477(8) days, the precession cycle is derived as 1.705(2) days.

\subsection{TY Psc}

TY Psc is a long known SU UMa-type dwarf nova with a V magnitude range from 17.5 to 12.2. The superhump period of which was firstly determined to be 0.07014 days by \citet{1988ApJ...334..422S}. Based on time-resolved spectroscopy, the orbital period of 0.06833(5) days was obtained from velocities of their $H\alpha$ emission lines \citep{1996PASP..108...73T}. The supercycle of this object is 210 days and the normal outburst cycle is 39 days \citep{2003A&A...404..301R}.  \citet{2009PASJ...61S.395K} observed the 2005 and 2008 superoutbursts partly, only part of stages B and C were observed during the 2008 superoutburst, and the superhump period of 0.07045(2) days was reported. 

\begin{table}[h]
	\centering
	\caption{Superhump period in TY Psc}
	\label{tab:typsc}
	\begin{tabular}{lccccc} 
		\hline
		sector & stage & E & BJD 2457000+  & period(d) & Error\\
		\hline
		
		17 &  A & 0 - 12  & 1769.55 - 1770.39  & 0.071037 &0.000061 \\
		&  B & 13 - 41 & 1770.39 - 1772.45 &0.070691&0.000020\\

		\hline
	\end{tabular}
\end{table}

\begin{figure}[h]
	\centering
	\includegraphics[width=0.6\columnwidth]{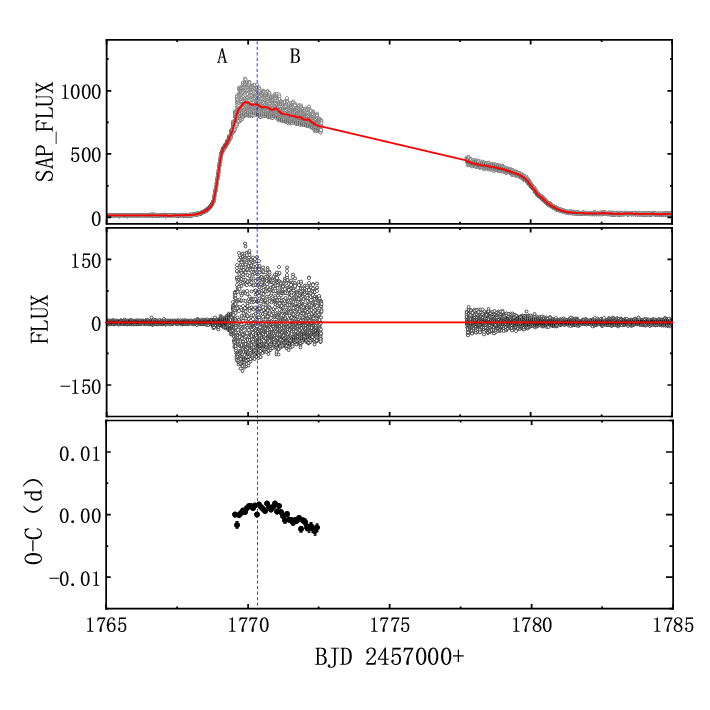}
	\caption{Top: the superoutburst light curves of TY Psc, the red line shows the trend of the superoutburst. Middle: the detrended light curves. Bottom: O-C diagram of TY Psc during BJD 2458768 -2458781 superoutburst. The ephemeris of $T_{max} = BJD 2458769.5520261+0.070756E$ was used for drawing this O-C diagram. The  blue dashed line represents the temporal dividing line between stage A and B. }
	\label{fig:typsc}
\end{figure}

The 2019 superoutburst of TY Psc was observed by TESS (see figure \ref{fig:typsc}), but a five-day data gap locates in decrease phase of the superoutburst. It lasts a total of 13 days from Barycentric Julian Date (BJD) 2458768 to 2458781. The O-C analysis shows that the stages A and B and a total of 41 cycles of superhumps were observed in this superoutburst. The transition of stage B to C and the whole stage C were not observed.  It is worth noting that the superhump period variation is tiny, which indicates that the period decrease rate is small.  The information of each stage superhumps are shown in table \ref{tab:typsc}. The average superhump period of this superoutburst is 0.070756(14) days. According to the stage A superhump period of 0.071037(61) days, the q = 0.106(5) is derived. Based on the average superhump period, the precession cycle is derived as 1.99(3) days.

\subsection{SS UMi}

SS UMi is considered to be a high mass-transfer rate dwarf nova. Its supercycle of 85 days close to the supercyle (20-60 days) of ER UMa stars \citep{2013arXiv1301.3202K}. The orbital period of 0.06778(4) days was obtained by time-resolved spectroscopy \citep{1996PASP..108...73T}. 
Using photometric observations, \citet{2013PASJ...65...23K} obtained a photometric orbital period of 0.067855(7) days. Based on the observations of a 2004 superoutburst, \citet{2006A&A...452..933O} reported a mean superhump period of 0.070149(16) days. They claimed that SS UMi lies in the transition area between normal SU UMa stars and ER UMa stars.   Recently study of its superhumps was given by \citet{2013PASJ...65...23K}, a 0.070358 days mean superhump period of a 2012 superoutburst was reported.
\begin{table}[h]
	\centering
	\caption{Superhump period in SS UMi}
	\label{tab:SSUMi}
	\begin{tabular}{lccccc} 
		\hline
		sector & stage & E & BJD 2457000+  & period(d) &Error\\
		\hline
		
		52  &  A & 0 - 23  & 2721.42 - 2723.05  & 0.070960 & 0.000034 \\
		&  B & 24 - 77 & 2723.05 - 2726.84 & 0.070159& 0.000014 \\
		&  C & 78 - 103 & 2726.84- 2728.65 & 0.069797& 0.000035 \\

		\hline
	\end{tabular}
\end{table}

Three superoutbursts were observed by TESS, but two of them were observed only very few parts. Therefore, we chose to analyze the fully observed superoutburst, which lasts 14 days from BJD 2459718 to 2459732.  The superhump ephemeris of $T_{max} = BJD 2459721.41611+0.070253E$ was used to calculated the O-C values. The O-C diagram shows typical 
three stages and a total of 103 cycles of superhumps are well observed in the superoutburst. The information of each stage superhumps was listed in table \ref{tab:SSUMi}. The mean superhump period is revised to 0.070253(9) days. Combining the adopted orbital period of 0.06778(4) days, the mass ratio is derived to be q = 0.129(3). Based on the mean superhump period, the precession cycle is derived as 1.93(3) days. 

\begin{figure}[h]
	\centering
	\includegraphics{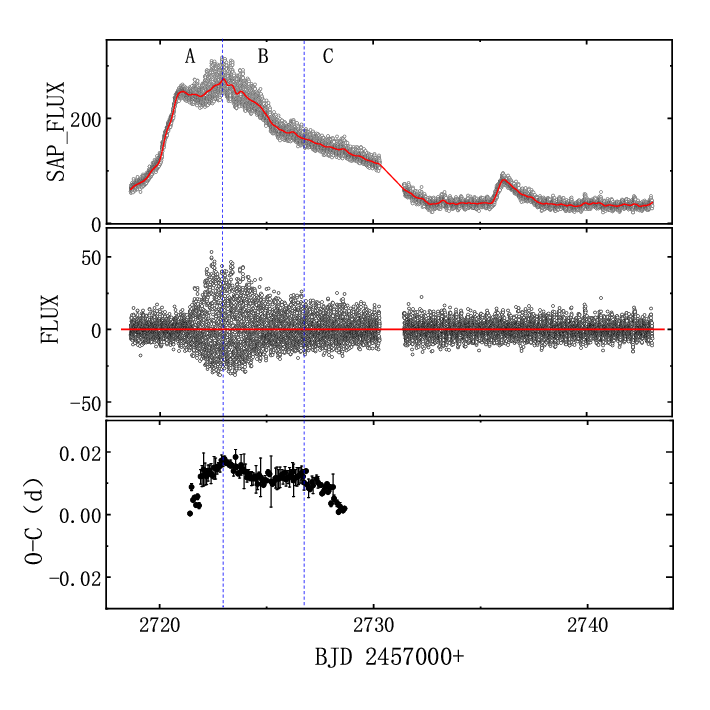}
	\caption{Top: the superoutburst light curves of SS UMi, the red line reprensents the trend of the superoutburst. Middle: the detrended light curves. Bottom: O-C diagram of SS UMi during BJD 2458768 -2458781 superoutburst. The ephemeris of $T_{max} = BJD 2459721.41611+0.070253E$ was used for drawing this O-C diagram. The blue dashed lines represent the temporal dividing line among stage A, B and C.  }
	\label{fig:SSUMi}
\end{figure}

\subsection{IX Dra}

IX Dra is a ER UMa-type star with a short supercycle of only 53 days and normal outburst cycle of 3-4 days which has a large duty cycle of superoutburst and a short outburst cycle \citep{2001PASJ...53L..51I,2003A&A...404..301R,2013MNRAS.429..868O}. ER UMa-type stars are thought to have higher mass transfer rate. \citet{2004AcA....54...57O} reported similar estimates of the supercycle and normal outburst cycle length based on the superoutburst in 2003. Additionally, they suggested the orbital period of 0.06646(6) days and suspected that this system is a period bouncer.  \citet{2007PASJ...59..929I} presented results of infrared spectroscopy to determine the spectral types of secondary stars. They indicated that the secondary is a more massive star. So according to \citet{2007PASJ...59..929I}, the probability that IX Dra is a period bouncer is slim.  Taking advantage of the superoutbursts of 2003 and 2012 July, \citet{2009PASJ...61S.395K,2014PASJ...66...90K} indicated that the superhumps can be expressed by a single period without a strong period variation. They also pointed out that the superhumps of IX Dra did not show any strong sign of a stage transition.  \citet{2013MNRAS.429..868O} proposed that the evaluation of the orbital period was problematic because they found two possible values, 0.06641(3) days ($95.70\pm 0.03$ min) and 0.06482(3) days ($93.34\pm 0.04$ min). Basing on 29 radial velocities in three night of 2014 June, \cite{2020AJ....160....6T} found $P_{orb} = 0.06480(16) $ days ($93.31\pm 0.23$ min) and argued that it is a pre-bounce system.  The orbital period of $0.06480(16) $ days was adopted in this work.  
\begin{table}[h]
	\centering
	\caption{Superhump period in IX Dra}
	\label{tab:IXDra}
	\begin{tabular}{lccccc} 
		\hline
		Sector & Stage & E & BJD 2457000+  & Period(d) & Error \\
		\hline
		16 &  A & 0 - 19   & 1740.36 - 1741.64& 0.067334 & 0.000042 \\
		&  B & 20 - 88  & 1741.64 - 1746.26& 0.067061 & 0.000012 \\
		&  C & 88 - 100 & 1746.26 - 1748.60& 0.066929 & 0.000067\\
		25 &  A & 0 - 22   & 1983.66 - 1985.14& 0.067170 & 0.000035\\
		&  B & 23 - 107 & 1985.14 - 1990.84& 0.067119 &0.000012\\
		&  C & 108 - 132& 1990.84 - 1992.50& 0.066677 & 0.000097 \\
		53 &  A & 0 - 16   & 2744.99 - 2746.07& 0.067215 &0.000050  \\
		&  B & 17 - 85  & 2746.07 - 2750.69& 0.067014 &0.000011 \\
		&  C & 86 - 104 & 2750.69 - 2751.96& 0.066976 &0.000087\\
		55 &  A & 0 - 18   & 2801.12 - 2802.33& 0.067268 & 0.000044 \\
		&  B & 19 - 87  & 2802.33 - 2806.88& 0.067030  & 0.000010\\
		&  C & 88 - 96  & 2806.88 - 2807.55& 0.066216 & 0.000233 \\
		\hline
	\end{tabular}
\end{table}

\begin{figure}[h]
	\centering
	
	\subfigure[$T_{max} = BJD 2458740.362971   + 0.067039E$]{
		\includegraphics[width=0.5\columnwidth]{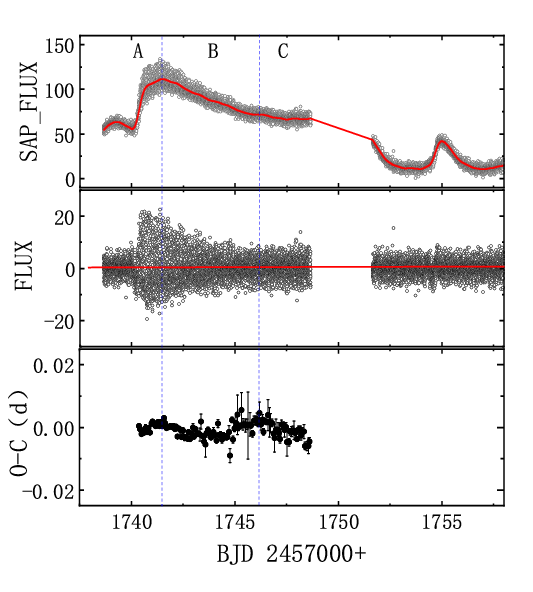}
	}%
	\subfigure[$T_{max} = BJD 245983.657722+ 0.067039E$]{
		\centering
		\includegraphics[width=0.5\columnwidth]{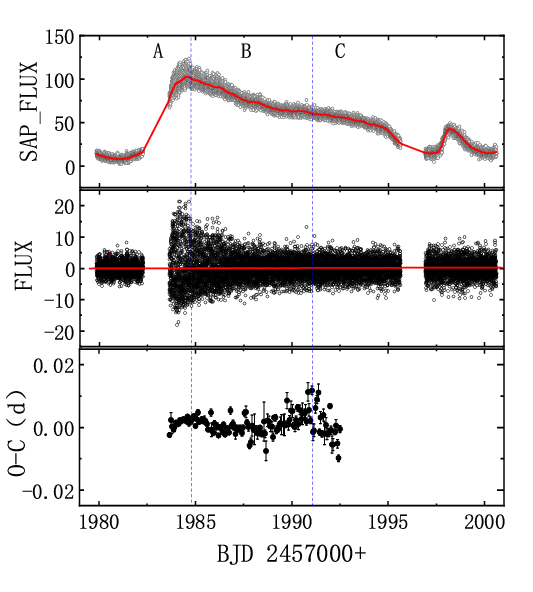}
	}
	
	\subfigure[$T_{max} = BJD 2459744.989897+ 0.067039E$]{
		\includegraphics[width=0.5\columnwidth]{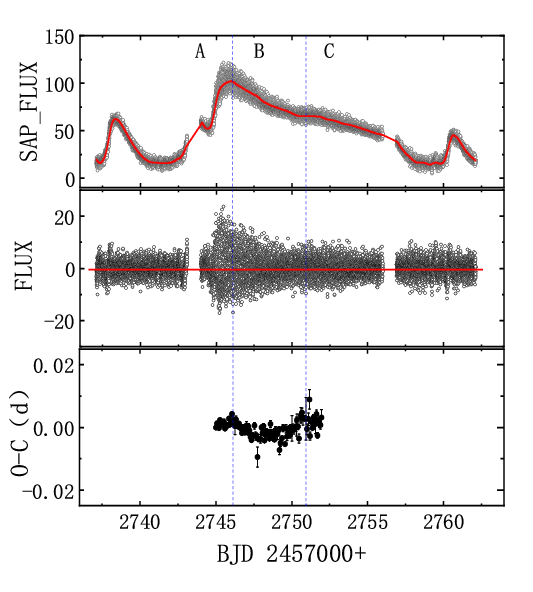}
	}%
	\subfigure[$T_{max} = BJD 2459801.116443+ 0.067039E$]{
		\includegraphics[width=0.5\columnwidth]{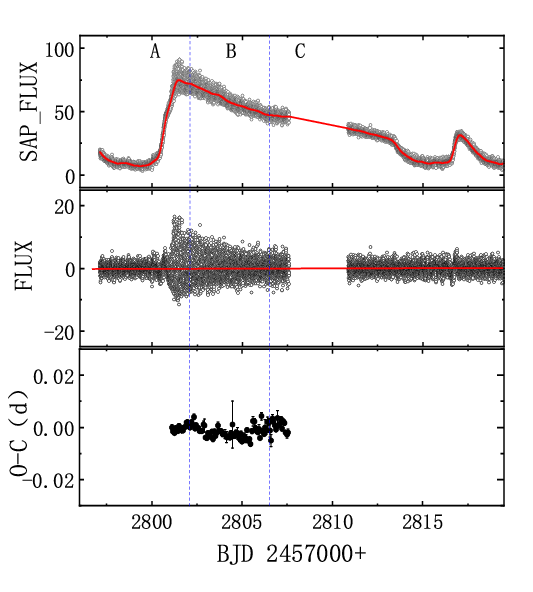}
	}%
	
	\centering
	\caption{ The four figures correspond to four superoutbursts of IX Dra. Top panel in each figure is light curves of superoutburst, the red lines show the trend of the superoutburst. Middle: the detrended light curves. Bottom: O-C curves of superhumps.  Ephemerides of Superhumps for each superoutburst are marked under each figure. The blue dashed lines represent the temporal dividing line among stage A, B and C.}
	\label{fig:IXDra}
\end{figure}

TESS has obtained eight superoutbursts of this object because the supercycle length is very short. We selected four well observed superoutbursts to analyse (see figure \ref{fig:IXDra}).  Although the superhump period variation is weak, but according to the four O-C results the stage transitons are clear.  It is worth noting that the change in stage B is slight and long-lasting. The information of each stage superhumps are shown in table \ref{tab:IXDra}. The superhump ephemerides are marked under each panel in figure \ref{fig:IXDra}. Mean superhump periods of the four superoutbursts are revised to 0.067057(8) days, 0.067053(8) days,0.067032(9) days and 0.067015(8) days. Based on the mean superhump period of 0.067039(8) days, the precession cycle is derived as 1.94(0.14) days.  The mass ratio of q = 0.101(9) is obtained based on the mean period [0.067247(43) days] of the four stage A superhump periods.

\subsection{DT Oct}
DT Oct (NSV 10934) shows very broad emission lines from the Far Ultraviolet Spectroscopic Explorer (FUSE) observations, which implies that the orbital inclination of this system is greater than 60° \citep{2009ApJ...701.1091G}.  However, eclipses have not been detected in its light curves.  Using the near-quiescent data in 2013, a possible orbital signal of 0.072707(5)d was reported by \citet{2014PASJ...66...90K}. The supercycle is unknown and the normal outburst cycle of 40-60 days which is relatively long \citep{2003A&A...404..301R}.
The superhumps has been studied by the observations of superoutbursts in 2003, 2008 and 2014\citep{2009PASJ...61S.395K,2014PASJ...66...90K,2015PASJ...67..105K}, and a mean superhump period of 0.07485 days was derived. 

\begin{table}[h]
	\centering
	\caption{Superhump period in DT Oct}
	\label{tab:DTOct}
	\begin{tabular}{lccccc} 
		\hline
		Sector & Stage & E & BJD 2457000+  & Period(d) & Error \\
		\hline
		27 &  A & 0 - 8   & 2058.55 - 2059.16& 0.076218 & 0.000130 \\
		&  B  & 9 - 22  & 2059.16 - 2060.20& 0.075095 & 0.000053 \\
		39 &  A & 0 - 11   & 2381.49 - 2382.33& 0.076858 & 0.000093\\
		&  B & 12 - 109& 2382.33 - 2389.65& 0.074812 & 0.000007 \\	
		\hline
	\end{tabular}
\end{table}

\begin{figure}[h]
	
	\subfigure[$T_{max} = BJD 2459058.549602+ 0.075120(5)E$]{
		\includegraphics{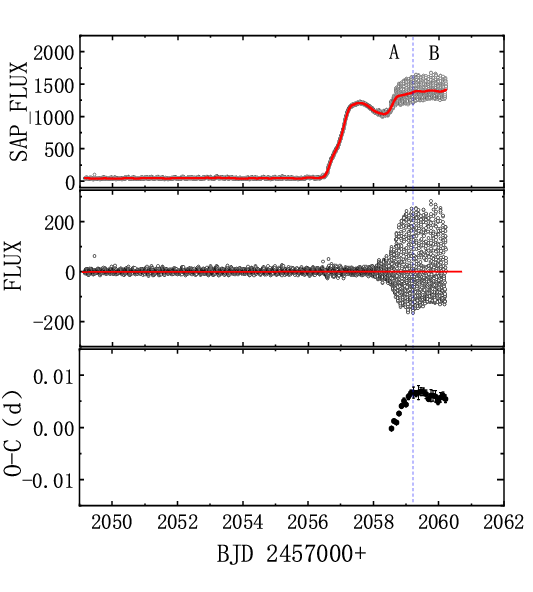} 
	}
	\subfigure[$T_{max} = BJD 2459381.4881014+ 0.075120(5)E$]{
		\includegraphics{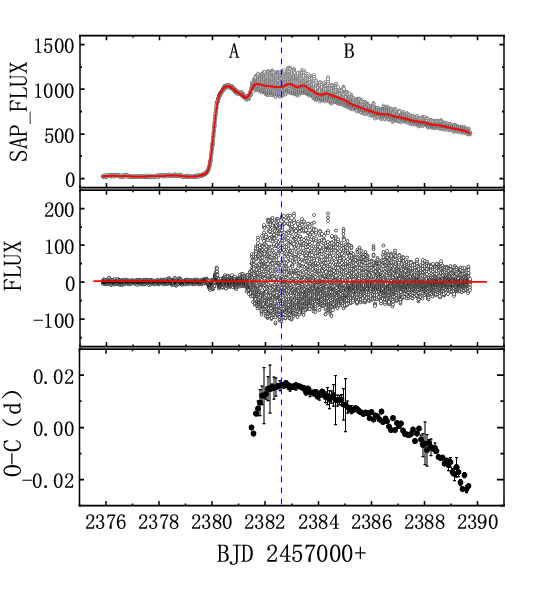} 
	}

	\caption{The two figures correspond to two superoutbursts of DT Oct. Top panels in each figure are light curves of superoutbursts, the red lines show the trend of the superoutbursts. Middle: the detrended light curves. Bottom: O-C curves of superhumps.  Ephemerides of superhumps for each superoutburst are marked under each figure. The blue dashed lines represent the temporal dividing line among stage A, B and C. }\label{fig:DTOct}
\end{figure}

TESS has obtained two superoutbursts of this star. The first superoutburst was observed only a few superhumps at the peak of superoutburst (see panel (a) in figure \ref{fig:DTOct}). Through the O-C analysis, it is found that the transition between stage B and C is not obvious in the second superoutburst (see panel (b) in figure \ref{fig:DTOct}) , So the stage C was not labeled in the figure. Meanwhile, the change in superhump period is very clear. We adopted the orbital period of 0.072707(5) days. Combined with the mean stage A superhump period of 0.076538(112) days, the resulting mass ratio is q = 0.147(5). This result is consistent with the results of \citet{2014PASJ...66...90K}, which indicates that this value is reliable. The mean superhump period is revised to 0.075120(7) days corresponds to precession cycle of 2.263(2) days. 

\subsection{J1730 +6247}
J1730 +6247 (SDSS J173008.38+624754.7)  was classified as a dwarf nova during the course of the Sloan Digital Sky Survey (SDSS) with a long supercycle of 327 days \citep{2002AJ....123..430S,2003A&A...404..301R}. The orbital period of J1730 +6247 is 110.22(12) minutes [0.076542(83) days], which was measured by SDSS spectroscopic data \citep{2009MNRAS.397.2170G,2015AJ....149..128T}. Its emission lines in the spectrum are single-peaked, suggesting a fairly low orbital inclination.  Based on the observations in 2001 and 2002 superoutbursts, \citet{2009PASJ...61S.395K} reported its mean superhump period was 0.0794 days. 

\begin{figure}[h]
	
	\subfigure[$T_{max} = BJD 2458757.338038+ 0.079664E$]{
		\includegraphics{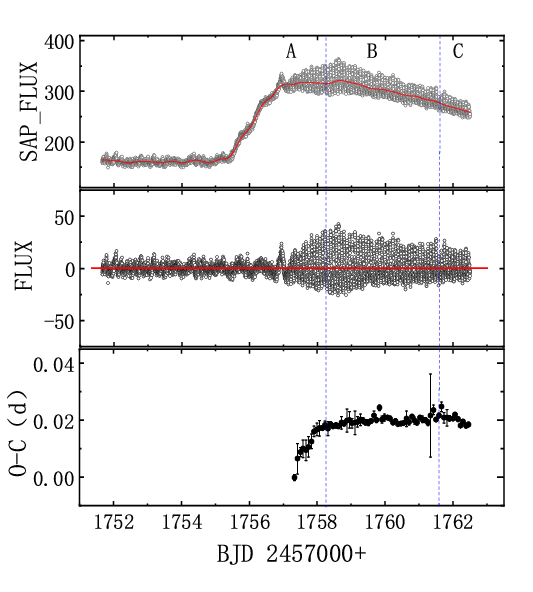} 
	}
	\subfigure[$T_{max} = BJD 2459912.7483359+ 0.079664E$]{
		\includegraphics{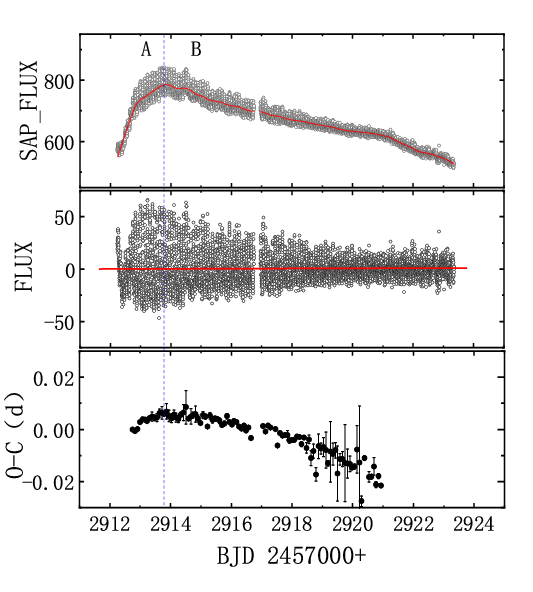} 
	}

	\caption{The two figures correspond to two superoutbursts of J1730 +6247. Top panels in each figure are light curves of superoutbursts, the red lines show the trend of the superoutbursts. Middle: the detrended light curves. Bottom: O-C curves of superhumps. Ephemerides of   superhumps for each superoutburst are marked under each figure. The blue dashed lines represent the temporal dividing line among stage A, B and C. }\label{fig:J1730+6247}
\end{figure}

Two superoutbursts of this target were obtained by TESS. The later stage of 2019 September-October superoutburst was not well observed (see panel (a) of figure \ref{fig:J1730+6247}). Three stages of superhumps in this superoutburst can still be distinguished in the O-C diagram.  
The O-C analysis of the fully observed 2022 November superoutburst  shows that the transition from stage B to C is not obvious (see panel (b) of figure \ref{fig:J1730+6247}).  The stage A superhump periods are determined to be 0.081007(96) and 0.080210(54) days (see table \ref{tab:J1730+6247}).
Combining the two stage A superhump periods, we get a mean period of 0.080609(75) days.  The mass ratio of q = 0.148(7) of this system is derived. The mean superhump period of the two superoutbursts is revised to 0.079664(7) days corresponds to a precession cycle of 1.95(5) days. 

\begin{table}[h]
	\centering
	\caption{Superhump period in J1730 +6247}
	\label{tab:J1730+6247}
	\begin{tabular}{lccccc} 
		\hline
		Sector & Stage & E & BJD 2457000+  & Period(d) & Error \\
		\hline
		17 &  A & 0 - 11   & 1757.34 - 1758.23 & 0.081007 & 0.000096 \\
		&  B  & 12 - 54    & 1758.23 - 1761.66 & 0.079749 & 0.000014 \\
		&  C  & 55 - 64    & 1761.66 - 1762.46 & 0.079625 & 0.000133 \\
		59 &  A & 0 - 14   & 2912.75 - 2913.87& 0.080210 & 0.000054\\
		&  B& 15 - 103& 2913.87 - 2920.93& 0.079422 & 0.000007 \\	
		\hline
	\end{tabular}
\end{table}

\begin{table}[h]
	\centering
	\caption{Results of the seven SU UMa stars}
	\label{tab:statistics}
	\begin{tabular}{lcccccccccccc} 
	\hline
	target & $P_{orb}(d)$ &Error& $\overline{P}_{sh}$(d) & Error &$P_{shA}$(d) &Error&$\epsilon^{\ast}$ &Error& q &Error & $P_{prec}$(d)&Error\\
	\hline
	V1504 Cyg &  0.06951 & 0.00005   & 0.072214  & 0.000008 & 0.073302 &0.000104&0.05173&0.00203&0.153&0.007&1.86&0.03 \\
	V503 Cyg &  0.077760 & 0.000003   & 0.081477  & 0.000008 & 0.082781 &0.000044&0.06065&0.00053&0.186&0.002&1.705&0.002 \\
	TY Psc &  0.06833 & 0.00005   & 0.070756  & 0.000014& 0.071037 &0.000061&0.03810&0.00152&0.107&0.005&1.99&0.03 \\
	SS UMi &  0.06778 & 0.00004   & 0.070253  & 0.000009& 0.070960 &0.000034&0.04481&0.00102&0.129&0.003&1.93&0.03 \\
	IX Dra  &  0.06646  & 0.00006   & 0.067039 & 0.000008 & 0.067247 & 0.000043&  0.03639 & 0.00299 & 0.101 & 0.009 & 1.94 & 0.14 \\
	DT Oct    & 0.072707 &0.000005  & 0.075120  & 0.000007& 0.076538 & 0.000112& 0.05005 & 0.00146 & 0.147&0.005&2.263&0.002\\
	J1730 +6247	& 0.076542  & 0.000083& 0.079664&0.000007&0.080609 & 0.000054 &	0.05045&0.00191&0.148&0.007&1.95&0.05 \\		
	\hline 
		
	\end{tabular}
	$P_{orb}$ : the orbital period;  $\overline{P}_{sh}$ : the mean superhump period; $P_{shA}$ : the stage A superhump period, if more than one stage A is observed, then the average is taken; $\epsilon^{\ast}$: the fractional superhump excess; q: mass ratio (M2/M1); $P_{prec}$: the precession period of accretion disk.

\end{table}

\begin{figure}[h]
	\centering
	\includegraphics{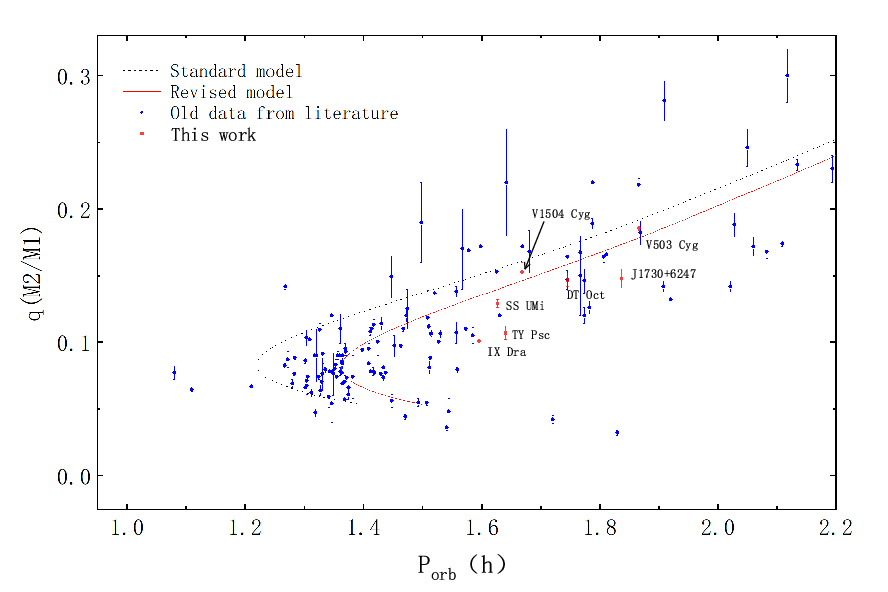}
	\caption{Mass ratio – orbital period relation of CVs. The data of blue points are form the literature \citep{2022arXiv220102945K}. The data of the red dots are obtained by this work. The black dashed line shows the standard model for CV evolution. The red solid line shows the evolution of donor properties along the revised model track \citep{2011ApJS..194...28K}. }
	\label{fig:pqt}
\end{figure}
\begin{figure}
	\centering
	\includegraphics{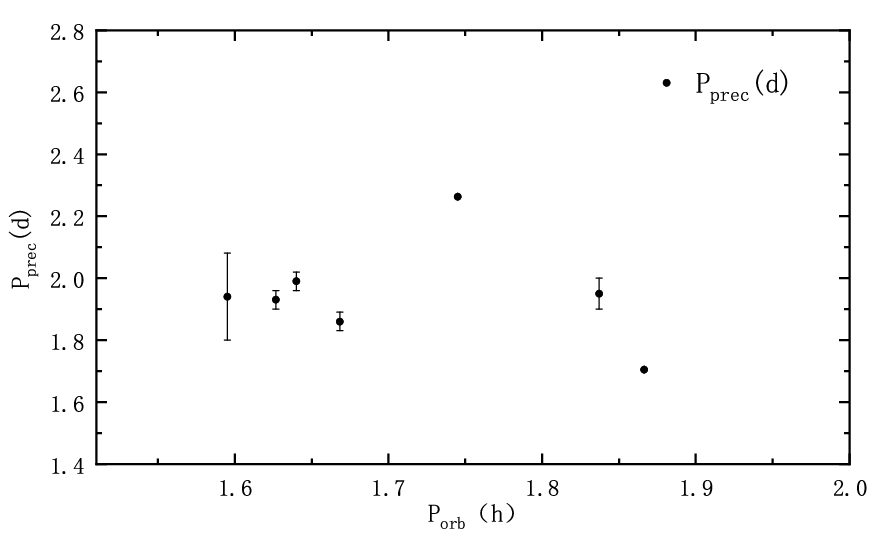}
	\caption{The distribution of precession periods with orbital periods. }
	\label{fig:prect}
\end{figure}

\section{Conclusion}

Using TESS observations, we analyzed the changes in the periods and amplitudes of superhumps in superoutbursts of the seven non-eclipsing SU UMa stars. We identified stages A, B, and C for each superoutburst by the O-C method and obtained the precise superhump periods for each stage. The information on each superhump stage of superoutbursts in the seven stars is listed in tables 2-8. The information in the tables shows that the mean superhump periods of stage A are longer than stage B and the mean superhump periods of B are longer than stage C, which is fully consistent with the stages A, B, and C superhump model.  Five out of the seven SU UMa stars in this work show clear three stages. All these seven stars show clear stage A and the transitions from stage A to B.  However, stage C was not determined in some superoutbursts. This is mainly due to observations do not fully cover the superoutbursts.  In addition, The transition from stage B to C is not obvious in some targets or superoutbursts (e.g., DT Oct and J1730 +6247). The two O-C diagrams for J1730 +6247 show that the transition from stage B to C can be distinguished in its first superoutburst but can not be distinguished in its second superoutburst (see figure \ref{fig:J1730+6247}). What causes this discrepancy is unknown. The O-C analysis shows that the variation rate of the superhump period is different for different systems.  Overall, the amplitude quickly increases in the plateau and then gradually decreases with time. The maximum amplitude moment corresponds to the light maximum of the superoutbursts.  The transition moments of the seven stars from stage A to B are about the moments of their superhump reaches maximum amplitude. 

The mass ratios (M2/M1) of these targets were obtained by using the stage A superhump method.  Mass ratio is an important parameter for binary stars, and it is not easy to obtain mass ratios of non-eclipsing systems. The precession periods were calculated by equation \ref{Eq:prec}. The results for the seven stars were summarized in table \ref{tab:statistics}.  We plotted the resulting mass ratios of this work and their orbital periods on the map of the evolution of CVs (see figure \ref{fig:pqt}). The figure shows that all the seven stars are pre-bounce systems with $q$ ranging from 0.1 to 0.2.  IX Dra was classified as a period bouncer by \citet{2004AcA....54...57O}.  The result shows that it is not a post-bounce system in agreement with the conclusion proposed by \citet{2020AJ....160....6T}.  There is no evidence shows that supercycle is associated with mass ratio. The mass ratio of TY Psc is 0.107(5) with the supercycle of 210 days.  SS UMi, which is a short supercycle system, has a larger mass ratio of 0.129(3) than TY Psc. The precession period of the seven SU UMa stars ranges from 1.7 to 2.3 days(see figure \ref{fig:prect}). The result does not show a strong linear correlation between the precession period and the orbital period.    Of the seven stars, V503 Cyg has the longest orbital period and largest mass ratio, but its precession cycle is the shortest.  
\\ \hspace*{\fill} \\

This work is supported by the National Natural Science Foundation of China (No.11933008 and No.12103084 ), the basic research project of Yunnan Province (Grant No.202301AT070352), and Natural Science Foundation of Anhui Province (2208085QA23). This paper includes data collected by the TESS mission, which are publicly available from the Mikulski Archive for Space Telescopes  (MAST). Funding for the TESS mission is provided by the NASA Science Mission Directorate. We are grateful to the referee for his/her valuable comments and suggestions, which have improved the manuscript greatly.

\bibliography{sample631}{}

\begin{thebibliography}{}
\expandafter\ifx\csname natexlab\endcsname\relax\def\natexlab#1{#1}\fi
\providecommand{\url}[1]{\href{#1}{#1}}
\providecommand{\dodoi}[1]{doi:~\href{http://doi.org/#1}{\nolinkurl{#1}}}
\providecommand{\doeprint}[1]{\href{http://ascl.net/#1}{\nolinkurl{http://ascl.net/#1}}}
\providecommand{\doarXiv}[1]{\href{https://arxiv.org/abs/#1}{\nolinkurl{https://arxiv.org/abs/#1}}}

\bibitem[{{Antonyuk} \& {Pavlenko}(2005)}]{2005ASPC..330..379A}
{Antonyuk}, O.~I., \& {Pavlenko}, E.~P. 2005, in Astronomical Society of the
  Pacific Conference Series, Vol. 330, The Astrophysics of Cataclysmic
  Variables and Related Objects, ed. J.~M. {Hameury} \& J.~P. {Lasota}, 379

\bibitem[{{Breger} {et~al.}(1999){Breger}, {Handler}, {Garrido}, {Audard},
  {Zima}, {Papar{\'o}}, {Beichbuchner}, {Li}, {Jiang}, {Liu}, {Zhou}, {Pikall},
  {Stankov}, {Guzik}, {Sperl}, {Krzesinski}, {Ogloza}, {Pajdosz}, {Zola},
  {Thomassen}, {Solheim}, {Serkowitsch}, {Reegen}, {Rumpf}, {Schmalwieser}, \&
  {Montgomery}}]{1999A&A...349..225B}
{Breger}, M., {Handler}, G., {Garrido}, R., {et~al.} 1999, \aap, 349, 225

\bibitem[{{Coyne} {et~al.}(2012){Coyne}, {Shenoy}, {MacLachlan}, {Lewis},
  {Dhuga}, {Eskandarian}, {Cobb}, {Maximon}, \& {Parke}}]{2012arXiv1206.6762C}
{Coyne}, R., {Shenoy}, A., {MacLachlan}, G., {et~al.} 2012, arXiv e-prints,
  arXiv:1206.6762, \dodoi{10.48550/arXiv.1206.6762}

\bibitem[{{G{\"a}nsicke} {et~al.}(2009){G{\"a}nsicke}, {Dillon}, {Southworth},
  {Thorstensen}, {Rodr{\'\i}guez-Gil}, {Aungwerojwit}, {Marsh}, {Szkody},
  {Barros}, {Casares}, {de Martino}, {Groot}, {Hakala}, {Kolb}, {Littlefair},
  {Mart{\'\i}nez-Pais}, {Nelemans}, \& {Schreiber}}]{2009MNRAS.397.2170G}
{G{\"a}nsicke}, B.~T., {Dillon}, M., {Southworth}, J., {et~al.} 2009, \mnras,
  397, 2170, \dodoi{10.1111/j.1365-2966.2009.15126.x}

\bibitem[{{Godon} {et~al.}(2009){Godon}, {Sion}, {Barrett}, \&
  {Szkody}}]{2009ApJ...701.1091G}
{Godon}, P., {Sion}, E.~M., {Barrett}, P.~E., \& {Szkody}, P. 2009, \apj, 701,
  1091, \dodoi{10.1088/0004-637X/701/2/1091}

\bibitem[{{Harvey} {et~al.}(1995){Harvey}, {Skillman}, {Patterson}, \&
  {Ringwald}}]{1995PASP..107..551H}
{Harvey}, D., {Skillman}, D.~R., {Patterson}, J., \& {Ringwald}, F.~A. 1995,
  \pasp, 107, 551, \dodoi{10.1086/133591}

\bibitem[{{Hellier}(2001)}]{2001cvs..book.....H}
{Hellier}, C. 2001, {Cataclysmic Variable Stars} (Springer Science \& Business
  Media)

\bibitem[{{Hirose} \& {Osaki}(1990)}]{1990PASJ...42..135H}
{Hirose}, M., \& {Osaki}, Y. 1990, \pasj, 42, 135

\bibitem[{{Howell} {et~al.}(2013){Howell}, {Everett}, {Seebode}, {Szkody},
  {Still}, {Wood}, {Ramsay}, {Cannizzo}, \& {Smale}}]{2013AJ....145..109H}
{Howell}, S.~B., {Everett}, M.~E., {Seebode}, S.~A., {et~al.} 2013, \aj, 145,
  109, \dodoi{10.1088/0004-6256/145/4/109}

\bibitem[{{Ishioka} {et~al.}(2001){Ishioka}, {Kato}, {Uemura}, {Iwamatsu},
  {Matsumoto}, {Martin}, {Billings}, \& {Novak}}]{2001PASJ...53L..51I}
{Ishioka}, R., {Kato}, T., {Uemura}, M., {et~al.} 2001, \pasj, 53, L51,
  \dodoi{10.1093/pasj/53.6.L51}

\bibitem[{{Ishioka} {et~al.}(2007){Ishioka}, {Sekiguchi}, \&
  {Maehara}}]{2007PASJ...59..929I}
{Ishioka}, R., {Sekiguchi}, K., \& {Maehara}, H. 2007, \pasj, 59, 929,
  \dodoi{10.1093/pasj/59.5.929}

\bibitem[{{Kato}(2022)}]{2022arXiv220102945K}
{Kato}, T. 2022, arXiv e-prints, arXiv:2201.02945,
  \dodoi{10.48550/arXiv.2201.02945}

\bibitem[{{Kato} {et~al.}(2002){Kato}, {Ishioka}, \&
  {Uemura}}]{2002PASJ...54.1029K}
{Kato}, T., {Ishioka}, R., \& {Uemura}, M. 2002, \pasj, 54, 1029,
  \dodoi{10.1093/pasj/54.6.1029}

\bibitem[{{Kato} {et~al.}(2013{\natexlab{a}}){Kato}, {Nogami}, {Baba},
  {Masuda}, {Matsumoto}, \& {Kunjaya}}]{2013arXiv1301.3202K}
{Kato}, T., {Nogami}, D., {Baba}, H., {et~al.} 2013{\natexlab{a}}, arXiv
  e-prints, arXiv:1301.3202, \dodoi{10.48550/arXiv.1301.3202}

\bibitem[{{Kato} \& {Osaki}(2013)}]{2013PASJ...65..115K}
{Kato}, T., \& {Osaki}, Y. 2013, \pasj, 65, 115, \dodoi{10.1093/pasj/65.6.115}

\bibitem[{{Kato} {et~al.}(2009){Kato}, {Imada}, {Uemura}, {Nogami}, {Maehara},
  {Ishioka}, {Baba}, {Matsumoto}, {Iwamatsu}, {Kubota}, {Sugiyasu}, {Soejima},
  {Moritani}, {Ohshima}, {Ohashi}, {Tanaka}, {Sasada}, {Arai}, {Nakajima},
  {Kiyota}, {Tanabe}, {Imamura}, {Kunitomi}, {Kunihiro}, {Taguchi}, {Koizumi},
  {Yamada}, {Nishi}, {Kida}, {Tanaka}, {Ueoka}, {Yasui}, {Maruoka}, {Henden},
  {Oksanen}, {Moilanen}, {Tikkanen}, {Aho}, {Monard}, {Itoh}, {Dubovsky},
  {Kudzej}, {Dancikova}, {Vanmunster}, {Pietz}, {Bolt}, {Boyd}, {Nelson},
  {Krajci}, {Cook}, {Torii}, {Starkey}, {Shears}, {Jensen}, {Masi}, {Hynek},
  {Nov{\'a}k}, {Koci{\'a}n}, {Kr{\'a}l}, {Ku{\v{c}}{\'a}kov{\'a}}, {Kolasa},
  {{\v{S}}tastn{\'y}}, {Staels}, {Miller}, {Sano}, {Ponthi{\`e}re},
  {Miyashita}, {Crawford}, {Brady}, {Santallo}, {Richards}, {Martin},
  {Buczynski}, {Richmond}, {Kern}, {Davis}, {Crabtree}, {Beaulieu}, {Davis},
  {Aggleton}, {Morelle}, {Pavlenko}, {Andreev}, {Baklanov}, {Koppelman},
  {Billings}, {Urban{\v{c}}ok}, {{\"O}gmen}, {Heathcote}, {Gomez}, {Voloshina},
  {Retter}, {Mularczyk}, {Z{\l}oczewski}, {Olech}, {Kedzierski}, {Pickard},
  {Stockdale}, {Virtanen}, {Morikawa}, {Hambsch}, {Garradd}, {Gualdoni},
  {Geary}, {Omodaka}, {Sakai}, {Michel}, {C{\'a}rdenas}, {Gazeas}, {Niarchos},
  {Yushchenko}, {Mallia}, {Fiaschi}, {Good}, {Walker}, {James}, {Douzu},
  {Julian}, {Butterworth}, {Shugarov}, {Volkov}, {Chochol}, {Katysheva},
  {Rosenbush}, {Khramtsova}, {Kehusmaa}, {Reszelski}, {Bedient}, {Liller},
  {Pojma{\'n}ski}, {Simonsen}, {Stubbings}, {Schmeer}, {Muyllaert}, {Kinnunen},
  {Poyner}, {Ripero}, \& {Kriebel}}]{2009PASJ...61S.395K}
{Kato}, T., {Imada}, A., {Uemura}, M., {et~al.} 2009, \pasj, 61, S395,
  \dodoi{10.1093/pasj/61.sp2.S395}

\bibitem[{{Kato} {et~al.}(2013{\natexlab{b}}){Kato}, {Hambsch}, {Maehara},
  {Masi}, {Miller}, {Noguchi}, {Akasaka}, {Aoki}, {Kobayashi}, {Matsumoto},
  {Nakagawa}, {Nakazato}, {Nomoto}, {Ogura}, {Ono}, {Taniuchi}, {Stein},
  {Henden}, {de Miguel}, {Dubovsky}, {Kudzej}, {Imamura}, {Akazawa}, {Takagi},
  {Wakabayashi}, {Ogi}, {Tanabe}, {Ulowetz}, {Morelle}, {Pickard}, {Ohshima},
  {Kasai}, {Pavlenko}, {Antonyuk}, {Baklanov}, {Antonyuk}, {Samsonov}, {Pit},
  {Sosnovskij}, {Littlefield}, {Sabo}, {Ruiz}, {Krajci}, {Dvorak}, {Oksanen},
  {Hirosawa}, {Goff}, {Monard}, {Shears}, {Boyd}, {Voloshina}, {Shugarov},
  {Chochol}, {Miyashita}, {Pietz}, {Katysheva}, {Itoh}, {Bolt}, {Andreev},
  {Parakhin}, {Malanushenko}, {Martinelli}, {Denisenko}, {Stockdale}, {Starr},
  {Simonsen}, {Tristram}, {Fukui}, {Tordai}, {Fidrich}, {Paxson}, {Itagaki},
  {Nakashima}, {Yoshida}, {Nishimura}, {Kryachko}, {Samokhvalov}, {Korotkiy},
  {Satovski}, {Stubbings}, {Poyner}, {Muyllaert}, {Gerke}, {MacDonald},
  {Linnolt}, {Maeda}, \& {Hautecler}}]{2013PASJ...65...23K}
{Kato}, T., {Hambsch}, F.-J., {Maehara}, H., {et~al.} 2013{\natexlab{b}},
  \pasj, 65, 23, \dodoi{10.1093/pasj/65.1.23}

\bibitem[{{Kato} {et~al.}(2014){Kato}, {Dubovsky}, {Kudzej}, {Hambsch},
  {Miller}, {Ohshima}, {Nakata}, {Kawabata}, {Nishino}, {Masumoto},
  {Mizoguchi}, {Yamanaka}, {Matsumoto}, {Sakai}, {Fukushima}, {Matsuura},
  {Bouno}, {Takenaka}, {Nakagawa}, {Noguchi}, {Iino}, {Pickard}, {Maeda},
  {Henden}, {Kasai}, {Kiyota}, {Akazawa}, {Imamura}, {de Miguel}, {Maehara},
  {Monard}, {Pavlenko}, {Antonyuk}, {Pit}, {Antonyuk}, {Baklanov}, {Ruiz},
  {Richmond}, {Oksanen}, {Harlingten}, {Shugarov}, {Chochol}, {Masi},
  {Nocentini}, {Schmeer}, {Bolt}, {Nelson}, {Ulowetz}, {Sabo}, {Goff}, {Stein},
  {Michel}, {Dvorak}, {Voloshina}, {Metlov}, {Katysheva}, {Neustroev},
  {Sjoberg}, {Littlefield}, {D{\k{e}}bski}, {Sowicka}, {Klimaszewski},
  {Cury{\l}o}, {Morelle}, {Curtis}, {Iwamatsu}, {Butterworth}, {Andreev},
  {Parakhin}, {Sklyanov}, {Shiokawa}, {Nov{\'a}k}, {Irsmambetova}, {Itoh},
  {Ito}, {Hirosawa}, {Denisenko}, {Kochanek}, {Shappee}, {Stanek}, {Prieto},
  {Itagaki}, {Stubbings}, {Ripero}, {Muyllaert}, \&
  {Poyner}}]{2014PASJ...66...90K}
{Kato}, T., {Dubovsky}, P.~A., {Kudzej}, I., {et~al.} 2014, \pasj, 66, 90,
  \dodoi{10.1093/pasj/psu072}

\bibitem[{{Kato} {et~al.}(2015){Kato}, {Hambsch}, {Dubovsky}, {Kudzej},
  {Monard}, {Miller}, {Itoh}, {Kiyota}, {Masumoto}, {Fukushima}, {Kinoshita},
  {Maeda}, {Mikami}, {Matsuda}, {Kojiguchi}, {Kawabata}, {Takenaka},
  {Matsumoto}, {de Miguel}, {Maeda}, {Ohshima}, {Isogai}, {Pickard}, {Henden},
  {Kafka}, {Akazawa}, {Otani}, {Ishibashi}, {Ogi}, {Tanabe}, {Imamura},
  {Stein}, {Kasai}, {Vanmunster}, {Starr}, {Oksanen}, {Pavlenko}, {Antonyuk},
  {Antonyuk}, {Sosnovskij}, {Pit}, {Babina}, {Sklyanov}, {Nov{\'a}k}, {Dvorak},
  {Michel}, {Masi}, {Littlefield}, {Ulowetz}, {Shugarov}, {Golysheva},
  {Chochol}, {Krushevska}, {Ruiz}, {Tordai}, {Morelle}, {Sabo}, {Maehara},
  {Richmond}, {Katysheva}, {Hirosawa}, {Goff}, {Dubois}, {Logie}, {Rau},
  {Voloshina}, {Andreev}, {Shiokawa}, {Neustroev}, {Sjoberg}, {Zharikov},
  {James}, {Bolt}, {Crawford}, {Buczynski}, {Cook}, {Kochanek}, {Shappee},
  {Stanek}, {Prieto}, {Denisenko}, {Nishimura}, {Mukai}, {Kaneko}, {Ueda},
  {Stubbings}, {Moriyama}, {Schmeer}, {Muyllaert}, {Shears}, {Modic}, \&
  {Paxson}}]{2015PASJ...67..105K}
{Kato}, T., {Hambsch}, F.-J., {Dubovsky}, P.~A., {et~al.} 2015, \pasj, 67, 105,
  \dodoi{10.1093/pasj/psv072}

\bibitem[{{Kato} {et~al.}(2016){Kato}, {Ishioka}, {Isogai}, {Kimura}, {Imada},
  {Miller}, {Masumoto}, {Nishino}, {Kojiguchi}, {Kawabata}, {Sakai}, {Sugiura},
  {Furukawa}, {Yamamura}, {Kobayashi}, {Matsumoto}, {Wang}, {Chou}, {Ngeow},
  {Chen}, {Panwar}, {Lin}, {Hsiao}, {Guo}, {Lin}, {Omarov}, {Kusakin},
  {Krugov}, {Starkey}, {Pavlenko}, {Antonyuk}, {Sosnjvskij}, {Antonyuk}, {Pit},
  {Baklanov}, {Babina}, {Itoh}, {Padovan}, {Akazawa}, {Kafka}, {de Miguel},
  {Pickard}, {Kiyota}, {Shugarov}, {Chochol}, {Krushevska},
  {Seker{\'a}{\v{s}}}, {Pikalova}, {Sabo}, {Dubovsky}, {Kudzej}, {Ulowetz},
  {Dvorak}, {Stone}, {Tordai}, {Dubois}, {Logie}, {Rau}, {Vanaverbeke},
  {Vanmunster}, {Oksanen}, {Maeda}, {Kasai}, {Katysheva}, {Morelle},
  {Neustroev}, \& {Sjoberg}}]{2016PASJ...68..107K}
{Kato}, T., {Ishioka}, R., {Isogai}, K., {et~al.} 2016, \pasj, 68, 107,
  \dodoi{10.1093/pasj/psw101}

\bibitem[{{Kato} {et~al.}(2020){Kato}, {Isogai}, {Wakamatsu}, {Hambsch},
  {Itoh}, {Tordai}, {Vanmunster}, {Dubovsky}, {Kudzej}, {Medulka}, {Kimura},
  {Ohnishi}, {Monard}, {Pavlenko}, {Antonyuk}, {Pit}, {Antonyuk}, {Babina},
  {Baklanov}, {Sosnovskij}, {Pickard}, {Miller}, {Maeda}, {de Miguel},
  {Brincat}, {Licchelli}, {Cook}, {Shugarov}, {Zaostrojnykh}, {Chochol},
  {Golysheva}, {Katysheva}, {Zubareva}, {Stone}, {Kasai}, {Starr},
  {Littlefield}, {Kiyota}, {Andreev}, {Sergeev}, {Ruiz}, {Myers}, {Simon},
  {Vasylenko}, {Sold{\'a}n}, {{\"O}gmen}, {Nakajima}, {Nelson}, {Masi},
  {Menzies}, {Sabo}, {Bolt}, {Dvorak}, {Stanek}, {Shields}, {Kochanek},
  {Holoien}, {Shappee}, {Prieto}, {Kojima}, {Nishimura}, {Kaneko}, {Fujikawa},
  {Stubbings}, {Muyllaert}, {Poyner}, {Moriyama}, {Maehara}, {Schmeer}, \&
  {Denisenko}}]{2020PASJ...72...14K}
{Kato}, T., {Isogai}, K., {Wakamatsu}, Y., {et~al.} 2020, \pasj, 72, 14,
  \dodoi{10.1093/pasj/psz134}

\bibitem[{{Knigge} {et~al.}(2011){Knigge}, {Baraffe}, \&
  {Patterson}}]{2011ApJS..194...28K}
{Knigge}, C., {Baraffe}, I., \& {Patterson}, J. 2011, \apjs, 194, 28,
  \dodoi{10.1088/0067-0049/194/2/28}

\bibitem[{{Lasota}(2001)}]{2001NewAR..45..449L}
{Lasota}, J.-P. 2001, \nar, 45, 449, \dodoi{10.1016/S1387-6473(01)00112-9}

\bibitem[{{Lomb}(1976)}]{1976Ap&SS..39..447L}
{Lomb}, N.~R. 1976, \apss, 39, 447, \dodoi{10.1007/BF00648343}

\bibitem[{{O'Donoghue}(2000)}]{2000NewAR..44...45O}
{O'Donoghue}, D. 2000, \nar, 44, 45, \dodoi{10.1016/S1387-6473(00)00012-9}

\bibitem[{{Olech} {et~al.}(2006){Olech}, {Mularczyk}, {Kedzierski},
  {Z{\l}oczewski}, {Wi{\'s}niewski}, \& {Szaruga}}]{2006A&A...452..933O}
{Olech}, A., {Mularczyk}, K., {Kedzierski}, P., {et~al.} 2006, \aap, 452, 933,
  \dodoi{10.1051/0004-6361:20054483}

\bibitem[{{Olech} {et~al.}(2004){Olech}, {Zloczewski}, {Mularczyk},
  {Kedzierski}, {Wisniewski}, \& {Stachowski}}]{2004AcA....54...57O}
{Olech}, A., {Zloczewski}, K., {Mularczyk}, K., {et~al.} 2004, \actaa, 54, 57,
  \dodoi{10.48550/arXiv.astro-ph/0403208}

\bibitem[{{Osaki}(1974)}]{1974PASJ...26..429O}
{Osaki}, Y. 1974, \pasj, 26, 429

\bibitem[{{Osaki}(1989)}]{1989PASJ...41.1005O}
---. 1989, \pasj, 41, 1005

\bibitem[{{Osaki} \& {Kato}(2013)}]{2013PASJ...65...50O}
{Osaki}, Y., \& {Kato}, T. 2013, \pasj, 65, 50, \dodoi{10.1093/pasj/65.3.50}

\bibitem[{{Otulakowska-Hypka} {et~al.}(2013){Otulakowska-Hypka}, {Olech}, {de
  Miguel}, {Rutkowski}, {Koff}, \& {B{\k{a}}kowska}}]{2013MNRAS.429..868O}
{Otulakowska-Hypka}, M., {Olech}, A., {de Miguel}, E., {et~al.} 2013, \mnras,
  429, 868, \dodoi{10.1093/mnras/sts385}

\bibitem[{{Pavlenko} {et~al.}(2012){Pavlenko}, {Samsonov}, {Antonyuk},
  {Andreev}, {Baklanov}, \& {Sosnovskij}}]{2012Ap.....55..494P}
{Pavlenko}, E.~P., {Samsonov}, D.~A., {Antonyuk}, O.~I., {et~al.} 2012,
  Astrophysics, 55, 494, \dodoi{10.1007/s10511-012-9255-4}

\bibitem[{{Press} \& {Rybicki}(1989)}]{1989ApJ...338..277P}
{Press}, W.~H., \& {Rybicki}, G.~B. 1989, \apj, 338, 277,
  \dodoi{10.1086/167197}

\bibitem[{Ricker {et~al.}(2015)Ricker, Winn, Vanderspek, Latham, Bakos, Bean,
  Berta-Thompson, Brown, Buchhave, Butler, {et~al.}}]{ricker2015transiting}
Ricker, G.~R., Winn, J.~N., Vanderspek, R., {et~al.} 2015, Journal of
  Astronomical Telescopes, Instruments, and Systems, 1, 014003,
  \dodoi{10.1117/1.JATIS.1.1.014003}

\bibitem[{{Ritter} \& {Kolb}(2003)}]{2003A&A...404..301R}
{Ritter}, H., \& {Kolb}, U. 2003, \aap, 404, 301,
  \dodoi{10.1051/0004-6361:20030330}

\bibitem[{{Scargle}(1982)}]{1982ApJ...263..835S}
{Scargle}, J.~D. 1982, \apj, 263, 835, \dodoi{10.1086/160554}

\bibitem[{{Szkody} \& {Feinswog}(1988)}]{1988ApJ...334..422S}
{Szkody}, P., \& {Feinswog}, L. 1988, \apj, 334, 422, \dodoi{10.1086/166846}

\bibitem[{{Szkody} {et~al.}(2002){Szkody}, {Anderson}, {Ag{\"u}eros},
  {Covarrubias}, {Bentz}, {Hawley}, {Margon}, {Voges}, {Henden}, {Knapp},
  {Vanden Berk}, {Rest}, {Miknaitis}, {Magnier}, {Brinkmann}, {Csabai},
  {Harvanek}, {Hindsley}, {Hennessy}, {Ivezic}, {Kleinman}, {Lamb}, {Long},
  {Newman}, {Neilsen}, {Nichol}, {Nitta}, {Schneider}, {Snedden}, \&
  {York}}]{2002AJ....123..430S}
{Szkody}, P., {Anderson}, S.~F., {Ag{\"u}eros}, M., {et~al.} 2002, \aj, 123,
  430, \dodoi{10.1086/324734}

\bibitem[{{Thorstensen}(2020)}]{2020AJ....160....6T}
{Thorstensen}, J.~R. 2020, \aj, 160, 6, \dodoi{10.3847/1538-3881/ab911c}

\bibitem[{{Thorstensen} {et~al.}(1996){Thorstensen}, {Patterson}, {Shambrook},
  \& {Thomas}}]{1996PASP..108...73T}
{Thorstensen}, J.~R., {Patterson}, J.~O., {Shambrook}, A., \& {Thomas}, G.
  1996, \pasp, 108, 73, \dodoi{10.1086/133693}

\bibitem[{{Thorstensen} \& {Taylor}(1997)}]{1997PASP..109.1359T}
{Thorstensen}, J.~R., \& {Taylor}, C.~J. 1997, \pasp, 109, 1359,
  \dodoi{10.1086/134016}

\bibitem[{{Thorstensen} {et~al.}(2015){Thorstensen}, {Taylor}, {Peters},
  {Skinner}, {Southworth}, \& {G{\"a}nsicke}}]{2015AJ....149..128T}
{Thorstensen}, J.~R., {Taylor}, C.~J., {Peters}, C.~S., {et~al.} 2015, \aj,
  149, 128, \dodoi{10.1088/0004-6256/149/4/128}

\bibitem[{{Vogt}(1974)}]{1974A&A....36..369V}
{Vogt}, N. 1974, \aap, 36, 369

\bibitem[{{Warner}(2003)}]{2003cvs..book.....W}
{Warner}, B. 2003, {Cataclysmic Variable Stars} (Cambridge University Press),
  \dodoi{10.1017/CBO9780511586491}

\end{thebibliography}
\bibliographystyle{aasjournal}



\end{document}